\documentclass[twocolumn,showpacs,amsmath,amssymb,prb]{revtex4}
\usepackage{graphicx}
\usepackage{textcomp}
\usepackage{nicefrac}
\usepackage[squaren]{SIunits}
\usepackage[utf8]{inputenc}

\begin{document}

\newcommand{\comment}[1]{!!! \textbf{#1} !!!}
\newcommand{\me}[0]{\mathrm{e}}
\newcommand{\mi}[0]{\mathrm{i}}
\newcommand{\md}[0]{\mathrm{d}}

\title{Interacting multi-component exciton gases in a potential trap: Phase separation and Bose--Einstein condensation}
\author{S.~Sobkowiak, D.~Semkat, H.~Stolz}
\affiliation{Institut f\"ur Physik, Universit\"at Rostock, 18051 Rostock, Germany}
\author{Th.~Koch, H.~Fehske}
\affiliation{Institut f\"ur Physik, Ernst--Moritz--Arndt--Universit\"at Greifswald, 17489 Greifswald, Germany}

\date{\today}

\begin{abstract}
The system under consideration is a multi-component gas of interacting paraexcitons and orthoexcitons confined in a three dimensional potential trap. We calculate the spatially resolved optical emission spectrum due to interband transitions involving weak direct and phonon mediated exciton-photon interactions.
For each component, the occurrence of a Bose-Einstein condensate changes the spectrum in a characteristic way so that it directly reflects the constant chemical potential of the excitons and the renormalization of the quasiparticle excitation spectrum. Moreover, the interaction between the components leads, in dependence on temperature and particle number, to modifications of the spectra indicating phase separation of the subsystems.
Typical examples of density profiles and luminescence spectra of ground-state paraexcitons and orthoexcitons in $\mathrm{Cu_2O}$ are given.
\end{abstract}



\maketitle

\section{Introduction}

Excitons in semiconductors have been promising candidates for the observation of Bose--Einstein condensation for several decades. At present, cuprous oxide (Cu$_2$O) is in the focus of experimental efforts due to the large binding energy and long lifetime of the exciton states. In order to obtain sufficiently high densities, entrapment by an external potential is an approved method.

The theoretical description of excitons in potential traps has been carried out so far mostly in the frame of a model of ideal bosons. Concepts for the inclusion of the interaction are well known from the theory of atomic condensates, \cite{griffin1996,DGP99,BFB00,PJ08} and first applications to excitons exist, too. \cite{BHB04} Recent investigations in the framework of a mean-field formalism in local density approximation have shown distinct signatures of a condensate in the decay luminescence spectrum of the non-condensed excitons. \cite{SS10} It is the aim of the present paper to introduce a generalization of this theory to a multi-component gas of interacting paraexcitons and orthoexcitons, where the consequences of the interaction on the condensation process are of particular interest. We show results for the densities of the individual components and their spatially resolved luminescence spectra for several parameter regimes and highlight experimentally relevant cases.

\section{Thermodynamics of excitons in a potential trap}

The thermodynamics of a one-component Bose gas has been investigated in detail, see, e.g., Refs.\ \onlinecite{griffin1996,DGP99,BFB00,PJ08}. First applications of these concepts to excitons have been presented in Ref.\ \onlinecite{BHB04} and, looking at spectral signatures of a condensate, in Ref.\ \onlinecite{SS10}. In analogy to generalizations for multi-component atomic gases, e.g., Refs.\ \onlinecite{HS96,BV97,SZC00,ZYY04,CQZ05,VS} and spinor polaritons, e.g., Refs.\ \onlinecite{SMK05,Ketal07}, in the following, we generalize this approach to the case of multiple species of excitons, i.e., paraexcitons and orthoexcitons, adopting a mean-field coupling scheme between the components. \cite{siegfrieddiplom}

The multi-component exciton gas is considered in second quantization. We start from the Hamiltonian for a $K$-component system in the grand canonical ensemble:
\begin{eqnarray}
\label{eq:Hamiltonallg}
\mathcal{H}&=& \sum_{i=1}^K \int \mathrm{d}\mathbf{r} \ \psi^\dagger_i(\mathbf{r},t)\left( -\frac{\hbar^2 \nabla^2}{2 M_i} + V_i(\mathbf{r})-\mu_i \right)  \psi_i(\mathbf{r},t) \nonumber \\
&+&\frac{1}{2}\sum_{i,j=1}^K \int \mathrm{d}\mathbf{r} \  h_{ij} \psi^\dagger_i(\mathbf{r},t) \psi^\dagger_j(\mathbf{r},t) \psi_j(\mathbf{r},t) \psi_i(\mathbf{r},t)\;,
\end{eqnarray}
with respective external potentials $V_i$ and chemical potentials $\mu_i$ for each species.
We assume a contact potential for the exciton--exciton interaction, with the matrix $h_{ij}$ containing the intra- and inter-species interaction strengths. Its components are given by the s-wave scattering lengths $a^s_{ij}$:
\begin{equation}
\label{eq:hij}
 h_{ij}=2\pi\hbar^2\left(\frac{1}{M_i}+\frac{1}{M_j}\right)a^s_{ij}\,.
\end{equation}
The Bose field operator $\psi_i$ obeys the Heisenberg equation of motion
\begin{eqnarray}
\label{eq:Heisenberg}
\mi\hbar\frac{\partial \psi_i(\mathbf{r},t)}{\partial t} &=& \left( -\frac{\hbar^2\nabla^2}{2M_i} +V_{i}(\mathbf{r})-\mu_i \right) \psi_i(\mathbf{r},t)\nonumber\\
&&+\sum_{j=1}^K h_{ij}\psi^\dagger_j(\mathbf{r},t) \psi_j(\mathbf{r},t) \psi_i(\mathbf{r},t)\,.
\end{eqnarray}
We decompose the field operators $\psi_i$ in the usual fashion,
\begin{equation}
\label{eq:Aufteilung}
\psi_i(\mathbf{r},t)=\varPhi_i(\mathbf{r})+\widetilde{\psi}_i(\mathbf{r},t)\,,
\end{equation}
where $\varPhi_i$ is the (scalar) condensate wave function with $\varPhi_i(\mathbf{r})=\langle \psi_i(\mathbf{r},t) \rangle=\langle \psi_i(\mathbf{r}) \rangle$ and $\widetilde{\psi}_i$ is the operator of the thermal excitons.
Inserting the decomposition (\ref{eq:Aufteilung}) into Eq.\ (\ref{eq:Heisenberg}) and following the steps of Ref.\ \onlinecite{griffin1996}, we obtain (arguments dropped for the sake of brevity)
\begin{eqnarray}
\label{eq:GPE3Kallgerg}
0\! &=&\! \Bigg(\! -\frac{\hbar^2\nabla^2}{2M_i} +V_i-\mu_i + h_{ii} \left ( n_{ii} + \widetilde n_{ii}\right ) +\!\sum_{j\neq i} h_{ij} n_{jj}  \!\Bigg) \varPhi_i \nonumber \\
&& + \, h_{ii} \widetilde{m}_{ii} \varPhi_i^* +\sum_{j\neq i} h_{ij} \Big(   \widetilde{n}_{ji} \varPhi_j  + \widetilde{m}_{ji}  \varPhi^*_j \Big)
\end{eqnarray}
and 
\begin{eqnarray}
\label{eq:Beweg3Kallgerg}
\mi\hbar\frac{\partial \widetilde{\psi}_i}{\partial t}\! &=&\! \ \Bigg(\! -\frac{\hbar^2\nabla^2}{2M_i} +V_i-\mu_i + 2 h_{ii}  n_{ii} +\!\sum_{j\neq i} h_{ij} n_{jj} \!\Bigg) \widetilde{\psi}_i\nonumber \\ 
&&+ \, h_{ii}m_{ii} \widetilde{\psi}^\dagger_i 
+ \sum_{j\neq i} h_{ij} \Big ( n_{ij} \widetilde{\psi}_j + m_{ij} \widetilde{\psi}^\dagger_j \Big )
\end{eqnarray}
with $n_{ij}\equiv \varPhi_j^\ast\varPhi_i + \widetilde n_{ij}$, $m_{ij}\equiv \varPhi_j\varPhi_i + \widetilde m_{ij}$, and the normal and anomalous averages $\widetilde{n}_{ij}=\langle \widetilde{\psi}^\dagger_i \widetilde{\psi}_j \rangle$ and $\widetilde{m}_{ij}=\langle \widetilde{\psi}_i \widetilde{\psi}_j \rangle$, respectively.
Equation (\ref{eq:GPE3Kallgerg}) generalizes the familiar Gross--Pitaevskii equation (GPE) by including (i) the coupling to the thermal excitons and (ii) the coupling of multiple components.

In a first approximation, we neglect all non-diagonal averages, i.e., $\widetilde{m}_{ij}=\widetilde{n}_{ij}=m_{ij}=n_{ij}=0 \quad\forall i\neq j$, reducing (\ref{eq:GPE3Kallgerg}) and (\ref{eq:Beweg3Kallgerg}) to effective one-species equations with an additional mean field contribution from the other species: 
\begin{eqnarray}
\label{eq:GPE3KallgergHFB}
0\! &=&\! \Bigg(\! -\frac{\hbar^2\nabla^2}{2M_i} +V_i-\mu_i + h_{ii} \left ( n_{ii} + \widetilde n_{ii}\right ) +\!\sum_{j\neq i} h_{ij} n_{jj} \!\Bigg) \varPhi_i \nonumber \\
&& + \, h_{ii} \widetilde{m}_{ii} \varPhi_i^* \,,
\end{eqnarray}
\begin{eqnarray}
\label{eq:Beweg3KallgergHFB}
\mi\hbar\frac{\partial \widetilde{\psi}_i}{\partial t}\! &=&\! \Bigg(\! -\frac{\hbar^2\nabla^2}{2M_i} +V_i-\mu_i + 2 h_{ii}  n_{ii} +\!\sum_{j\neq i} h_{ij} n_{jj} \!\Bigg) \widetilde{\psi}_i\nonumber \\ 
&&+ \, h_{ii}m_{ii} \widetilde{\psi}^\dagger_i\,.
\end{eqnarray}
Thanks to this simplification, Eq.\ (\ref{eq:Beweg3KallgergHFB}) can be formally solved by a Bogoliubov transformation,
\begin{equation}
\widetilde{\psi}_i=\sum\limits_\sigma\left[u_i(\sigma) a_i^{}(\sigma)\mathrm{e}^{-\mi E_i(\sigma) t/\hbar}+v_i^*(\sigma) a_i^{\dag}(\sigma) \mathrm{e}^{\mi E_i(\sigma) t/\hbar}\right]\,,
\end{equation} 
where $\sigma$ enumerates the quasiparticle states. Thereby the Bogoliubov amplitudes $u_i$ and $v_i$ satisfy the relation $\sum_\sigma [u_{i}(\sigma)^2 - v_{i}(\sigma)^2]=1$. The excitation spectrum $E_i(\sigma)$ is given by the solution of the eigenvalue problem
\begin{equation}
\label{eq:Bogomatrix}
\begin{pmatrix} \mathcal{L}_i & h_{ii}m_{ii} \\ -h_{ii}m_{ii}^{*} & -\mathcal{L}_i   \end{pmatrix} \begin{pmatrix} u_{i}(\sigma) \\ v_{i}(\sigma) \end{pmatrix} = E_{i}(\sigma) \begin{pmatrix} u_{i}(\sigma) \\ v_{i}(\sigma) \end{pmatrix} \,,
\end{equation}
with
\begin{equation}
\label{eq:defL3k}
\mathcal{L}_i = -\frac{\hbar^2\nabla^2}{2M_i} +V_{i}-\mu_i + 2 h_{ii} n_{ii} + \sum_{j\neq i} h_{ij} n_{jj}\,.
\end{equation}

Equations (\ref{eq:GPE3KallgergHFB}) and (\ref{eq:Bogomatrix}) are the multi-component generalizations of the Hartree--Fock--Bogoliubov (HFB) equations. They represent a system of $3K$ equations which are coupled via the interaction matrix elements $h_{ij}$.

To guarantee gapless spectra, we next apply the Popov approximation, i.e., we neglect the anomalous averages $\widetilde{m}_{ii}$ in (\ref{eq:GPE3KallgergHFB}) and (\ref{eq:Bogomatrix}) and find
\begin{equation}
\label{eq:GPE3KallgergHFBP}
0\!=\!\Bigg(\! -\frac{\hbar^2\nabla^2}{2M_i} +V_{i}-\mu_i + h_{ii}  \left ( n_{ii} + \widetilde n_{ii}\right )+\!\sum_{j\neq i} h_{ij} n_{jj}\!\Bigg) \varPhi_i 
\end{equation}
and
\begin{equation}
\label{eq:BogomatrixP}
\begin{pmatrix} \mathcal{L}_i & h_{ii}\varPhi_{i}^2 \\ -h_{ii}\varPhi_{i}^{*2} & -\mathcal{L}_i   \end{pmatrix} \begin{pmatrix} u_{i}(\sigma) \\ v_{i}(\sigma) \end{pmatrix} = E_{i}(\sigma) \begin{pmatrix} u_{i}(\sigma) \\ v_{i}(\sigma) \end{pmatrix} \,.
\end{equation}

Since the extension of the potential trap is large compared to the typical length scale of the system (e.g., the thermal deBroglie wavelength of the excitons), we can use the local density approximation (LDA). Then the excitons are treated as a locally homogeneous system and the spatial dependence enters only via the trap potential. In that case, the Bogoliubov equations are readily solved, yielding the density $n_i^T\equiv\widetilde n_{ii}$ of thermally excited excitons as
\begin{eqnarray}
\label{eq:nT3K}
n_{i}^T(\mathbf{r}) &=& \int \frac{\md^3\mathbf{k}}{8\pi^3} \left[ \frac{L_i(\mathbf{k},\mathbf{r})}{E_i(\mathbf{k},\mathbf{r})}\left( n_{B}(E_i(\mathbf{k},\mathbf{r}))+\frac{1}{2} \right)-\frac{1}{2} \right]\nonumber\\ &&\times\Theta\left(E_i(\mathbf{k},\mathbf{r})^2\right) 
\end{eqnarray}
with $n_B(E) = [exp(E/k_B T)-1]^{-1}$ being the usual Bose function. The excitation spectrum $E_i$ is explicitly given by
\begin{eqnarray}
\label{eq:Energie3k}
E_i(\mathbf{k},\mathbf{r})\! &=&\! \sqrt{L_i(\mathbf{k},\mathbf{r})^2-(h_{ii}n_{i}^c(\mathbf{r}))^2}\,, \\
L_i(\mathbf{k},\mathbf{r})\! &=&\! \frac{\hbar^2k^2}{2M_i} +V_{i}(\mathbf{r})-\mu_i + 2 h_{ii} n_{i}(\mathbf{r}) +\! \sum_{j\neq i} h_{ij} n_{j}(\mathbf{r})\nonumber\,,
\\
\end{eqnarray}
with $n_i^c\equiv|\varPhi_i|^2$ and $n_i\equiv n_{ii}=n_i^T+n_i^c$. 
In consistence with the LDA, we apply the Thomas--Fermi approximation to the Gross--Pitaevskii equation, neglecting the kinetic energy term in (\ref{eq:GPE3KallgergHFBP}). Then we obtain finally for the densities of the condensates 
\begin{eqnarray}
\label{eq:nc3K}
n_{i}^c(\mathbf{r})=\frac{1}{h_{ii}}\Bigg(\mu_i-V_{i}(\mathbf{r})
-2h_{ii}n_{i}^T(\mathbf{r})
-\sum\limits_{j\neq i} h_{ij} n_{j}(\mathbf{r})\Bigg)\nonumber\\
\times\Theta\Bigg(\mu_i-V_{i}(\mathbf{r})-2h_{ii}n_{i}^T(\mathbf{r})
-\sum_{j\neq i} h_{ij} n_{j}(\mathbf{r})\Bigg).\,
\end{eqnarray}
The expressions (\ref{eq:nT3K})--(\ref{eq:nc3K}) have to be solved self-consistently. Although they look very similar to the one-component case, a coupling between the components appears via $L_i$ and $n_{i}^c$.

In what follows we calculate the densities of excitons in Cu$_2$O in a strain induced potential trap. \cite{SN00} In addition to the paraexcitons (labeled `p'), two spin projections of orthoexcitons are captured by the trap, denoted by `+' and `$-$', while the zero component is expelled and plays no role. Thus, the number of components $K=3$. 
In addition to the usual symmetry of the interaction matrix, $h_{ij}=h_{ji}$ with $i,j=p,+,-$, in our case it holds that $h_{++}=h_{--}$ and $h_{p+}=h_{p-}$, leaving four independent parameters $h_{pp}$, $h_{++}$, $h_{p+}$, and $h_{+-}$. As extensive works on two-component systems \cite{HS96,BV97,SZC00,VS} have shown, one of the most interesting aspects of multicomponent systems---the occurrence of phase separation---is closely tied to the proportions of inter- and intra-species interaction strengths.

According to Eq.\ (\ref{eq:hij}), the interaction strengths are given by the s-wave
scattering lengths of the corresponding channels, which can be obtained by
the solution of the four-particle scattering problem. The case of
positronium--positronium scattering some time ago received much attention
\cite{SC01,Oda,Ivanov} and quite reliable values of the scattering length
for both the singlet and the triplet channel have been obtained. In
contrast, the description of exciton--exciton interaction is a long-standing
problem, and so far no satisfying solution for the general case has been
obtained. Especially for Cu$_2$O, we expect a strong effect of the
non-parabolicity of the valence band \cite{French} and of the rather large
electron--hole exchange interaction \cite{delta} on the scattering lengths.
Therefore, the values we deduced from the scattering lengths of the
positronium problem given in Ref.\ \onlinecite{SC01} ($h_{++}=0.71\, h_{pp}$, $h_{p+}=0.33\, h_{pp}$ and $h_{+-}=1.77\, h_{pp}$ with $h_{pp}=7.5\times 10^{-4}$ \textmu eV \textmu m$^3$), should be considered as representative, only. Nevertheless,
they allow to show the general behaviour of the multi-component exciton
system. In the calculation we also neglect the difference in the paraexciton and
orthoexciton mass due to the $k$-dependent exchange interaction. \cite{Dasbach}
\section{Luminescence spectrum}
Excitons decay by emitting photons. This takes place either directly, whereby momentum conservation requires that only excitons with the same momentum as the emitted photons are involved, or with assistance of momentum supplying phonons such that all exciton states can participate in the optical emission. Because the optical wavelength of the emission is much smaller than the trapped exciton cloud, we apply a local approximation to the emission spectrum, which, for the homogeneous case, is determined by the excitonic spectral function $A(\mathbf{k},\omega)$ \cite{shi1994,haug1983}:
\begin{eqnarray}
\label{eqn:spectrum}
I_i(\mathbf{r},\omega) & \propto & 2 \pi |S_i(\mathbf{k}=0)|^2 \delta(\hbar\omega^\prime - \mu_i) n_{i}^c(\mathbf{r}) \\
&& + \sum_{\mathbf{k}\ne 0}|S_i(\mathbf{k})|^2n_{B}(\hbar\omega^\prime - \mu_i)A_i(\mathbf{r},\mathbf{k},\hbar\omega^\prime - \mu_i) \nonumber
\end{eqnarray}
with $S_i(\mathbf{k})$ representing the exciton-photon coupling. The spectral function is given by the Bogoliubov amplitudes $u_i$ and $v_i$ and the quasiparticle spectrum in (\ref{eq:Energie3k}):
\begin{eqnarray}
A_i(\mathbf{r},\mathbf{k},\omega)&=&2\pi\hbar\Bigg[ u_i^2(\mathbf{k},\mathbf{r}) \delta(\hbar\omega-E_i(\mathbf{k},\mathbf{r})) \\
&&-\,v_i^2(\mathbf{k},\mathbf{r}) \delta(\hbar\omega+E_i(\mathbf{k},\mathbf{r})) \Bigg] \,.\nonumber
\end{eqnarray}
In order to account for a finite spectral resolution being important for 
comparison with measured spectra, we convolute the spectral intensity 
(\ref{eqn:spectrum}) with a slit function of the shape 
exp$[-(\omega/\Delta)^{4}]$. Here, $\Delta$ is a measure for the spectral 
resolution. Furthermore, in a typical experimental situation, one images a 
small stripe of width $2\Delta x$  elongated along the $z$-direction onto the 
entrance slit of a spectrograph. Integrating over the $y$-direction 
perpendicular to $z$ we obtain the spatially resolved spectrum of the thermal excitons (in the following, the direct condensate contribution is not considered):
\begin{eqnarray}
\label{eq:fertigallg}
I_i(\omega,z) &\propto& \int \limits_{-\Delta x}^{\Delta x } \md x \int 
\limits_{-\infty}^{\infty} \md y \int \md {\mathbf{k}} \, |S_i(\mathbf{k})|^2\,u_i^2
(\mathbf{k},\mathbf{r})
\nonumber\\
&&\hspace*{2ex}\times n_B(E_i(\mathbf{k},\mathbf{r})) \exp{[-\varepsilon_{-}^4 (\omega ',\mathbf{k},
\mathbf{r})]}
\nonumber \\
&&\hspace{-0.5cm} -\int \limits_{-\Delta x}^{\Delta x }\md x \int \limits_{-\infty}^{\infty} \md y 
\int \md {\mathbf{k}} \, |S_i(\mathbf{k})|^2\,v_i^2(\mathbf{k},\mathbf{r}) 
\nonumber \\
&&\hspace*{2ex}\times n_B(-E_i(\mathbf{k},\mathbf{r}))\exp{[ -\varepsilon_{+}^4 (\omega ',\mathbf{k},
\mathbf{r})]}
\end{eqnarray}
with $\varepsilon_{\pm} (\omega ',\mathbf{k},\mathbf{r})\equiv(\hbar 
\omega ' - \mu \pm E(\mathbf{k},\mathbf{r}))/\Delta$.

In the case of phonon-assisted transitions (being relevant for the orthoexcitons), we have $\omega^\prime = \omega - E_{gX}/\hbar -\omega_{\rm phonon}$ with $E_{gX}$ being the excitonic band gap of the semiconductor. We assume $S(\mathbf{k})$ to be a constant. Then the first term in Eq.\ (\ref{eqn:spectrum}) gives rise to a $\delta$-shaped luminescence line at the position of the chemical potential, with a strength determined by the condensate density.

For trapped paraexcitons, the zero-phonon decay is relevant and can be treated by setting $\omega^\prime = \omega - E_{gX}/\hbar$ and $S(\mathbf{k})= S_0 \delta(\mathbf{k}-\mathbf{k}_0)$. 
Here $\mathbf{k}_0$ is the wavevector of the intersection of photon and exciton dispersion. Its modulus is given by $k_0= E_{gX} n/\hbar c$, where $n$ is the refraction index and $c$ is the vacuum velocity of light. Due to the form of $S({\bf k})$, the condensate itself does not contribute to the direct luminescence process. 

However, as discussed for the one-component exciton gas in Ref.\ \onlinecite{SS10}, in case of a potential trap there will be indirect signatures of the condensate in the spatially resolved luminescence spectrum. The spectral lineshape follows the density distribution in the trap, which in turn is bordered by the minimal excitation energy $E({\bf k}=0,\mathbf{r})$. For a noncondensed gas the latter quantity is roughly parabolic due to the trapping potential, while it is zero in the presence of a condensate. Thus with increasing particle number (or decreasing temperature) the flat bottom of the spectrum of thermal excitons may be a footprint of Bose-Einstein condensation.

\section{Results}

We evaluate the density distributions of the trapped excitons in an 
iterative way. In each step, we keep the distributions of two of the 
components fixed. Under the constraint of fixed particle number, we 
iterate the subset of equations (\ref{eq:nT3K})--(\ref{eq:nc3K}) 
belonging to the third component to self-consistency. We cycle through 
the components until self-consistency of the whole system 
(\ref{eq:nT3K})--(\ref{eq:nc3K}) is achieved.
\begin{figure}[h]
\begin{minipage}[t]{\linewidth}
\includegraphics[width=0.95\linewidth,bb=-190 65 739 770, clip=true]{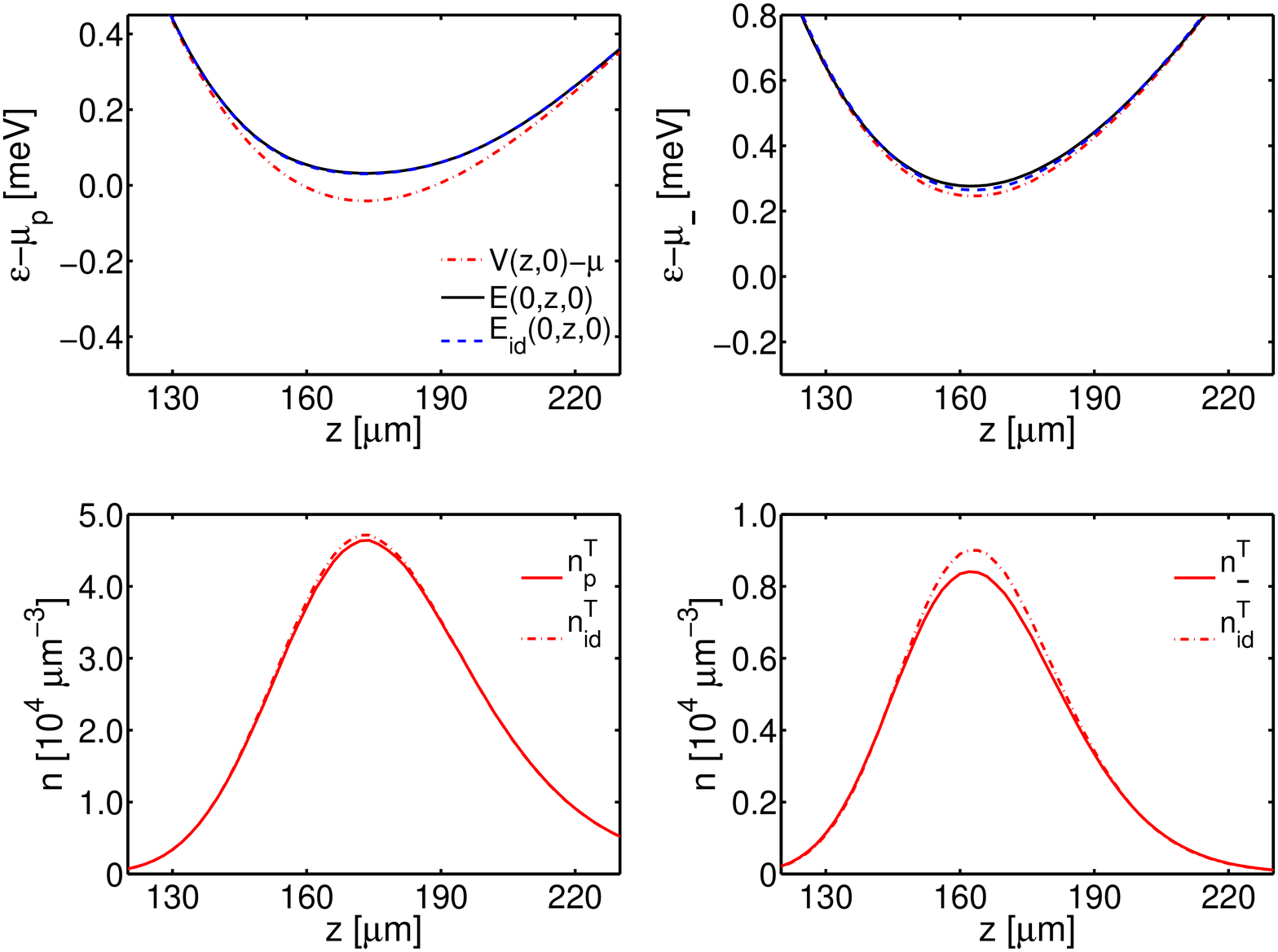}
\includegraphics[width=0.95\linewidth,bb=-190 65 739 770, clip=true]{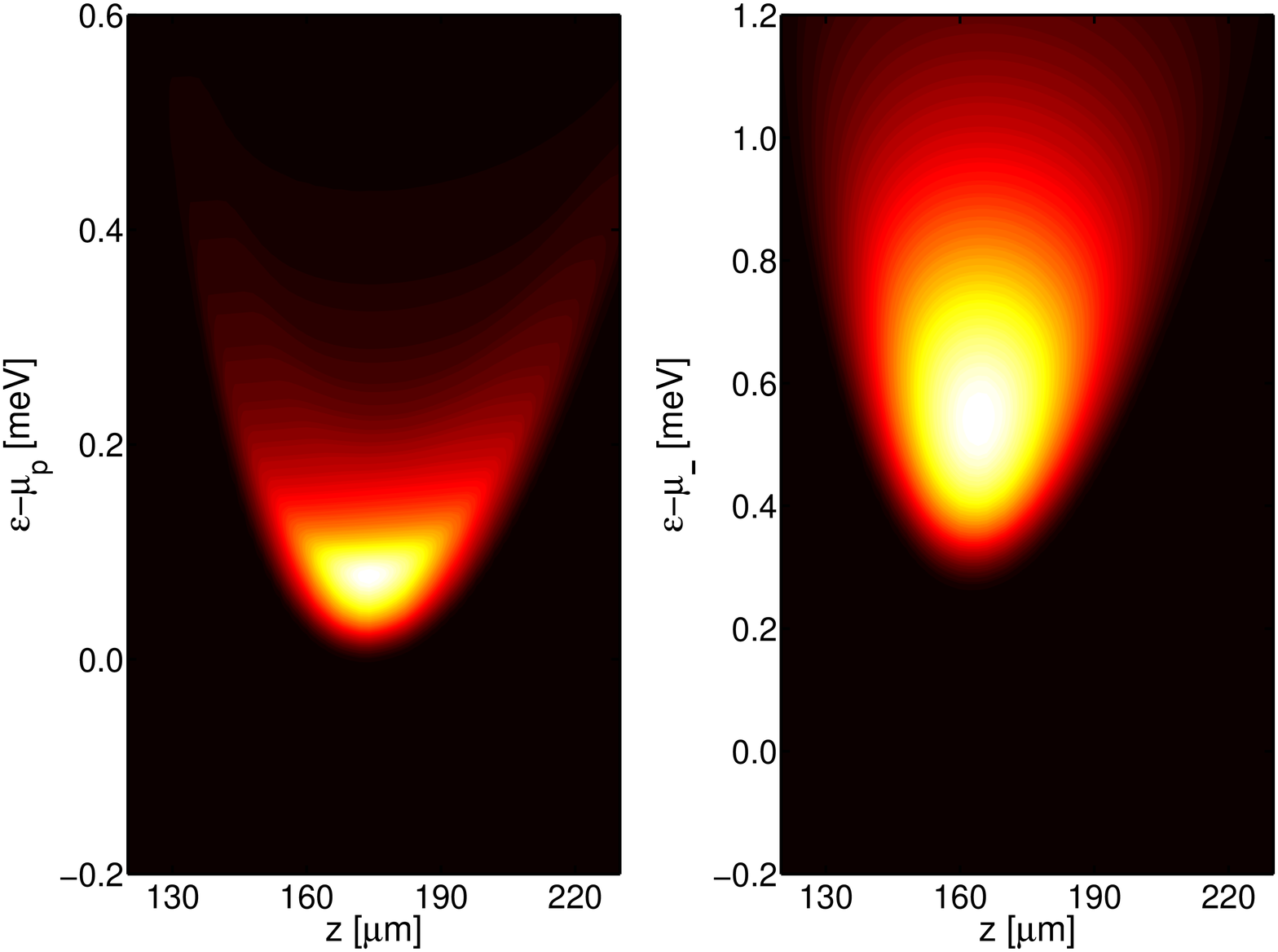}
\caption{Potentials and density profiles in $z$-direction at $(x,y)=(0,0)$, and luminescence spectra for paraexcitons (left column) and ortho($-$)excitons (right column) for a temperature of $T=\unit{2.2}{\kelvin}$ and particle numbers of $N_p=5 \times 10^9$ and $N_-=N_{+}=5 \times 10^{8}$ in the trap. The corresponding chemical potentials are $\mu_p=\unit{-2260}{\micro\electronvolt}$ and $\mu_{-}=\mu_{+}=\unit{-5920}{\micro\electronvolt}$. Upper row: external trap potential $V_i$, quasiparticle energy at $\mathbf{k}=0$ shifted by $\mu$ (i.e., renormalized potential) $E(0,z,0)$, and the same quantity without interspecies interaction $E_{id}(0,z,0)$. Middle row: densities of thermal excitons with ($n^T$) and without interspecies interaction ($n_{id}^T$). Lower row: luminescence spectra.}
 \label{fig1}
\end{minipage}
\end{figure}

Depending on the temperature and particle numbers in the trap, there may occur six distinct situations, featuring a condensate of (i) none of the species, (ii) only the paraexcitons, (iii) only one species of orthoexcitons, (iv) both species of orthoexcitons, (v) paraexcitons and one species of orthoexcitons, and (vi) all species. 
To analyze these cases, we set the particle numbers of each component to one of two values: $N_i=5\times 10^9$ or $N_i=5\times 10^8$ for $i=p,+,-$, respectively. We get a rough estimate of the corresponding critical temperatures by applying a harmonic approximation to the Hertzian potentials. Then, a simple Thomas-Fermi calculation for the single component case \cite{PS03} yields $T_c^0\approx \unit{2}{\kelvin}$ for $N=5\times 10^9$ and $T_c^0\approx \unit{1}{\kelvin}$ for $N=5\times 10^8$.

For our calculations, we use values of $\Delta=41$ \textmu eV for the spectral resolution and $\Delta x=25$ \textmu m for the entrance slit of the spectrograph being typical for a triple high resolution spectrograph used in the current experiments which are under way. \cite{schwartz2010}
\begin{figure}[h]
\begin{minipage}[t]{\linewidth}
\includegraphics[width=0.95\linewidth,bb=-190 65 739 770, clip=true]{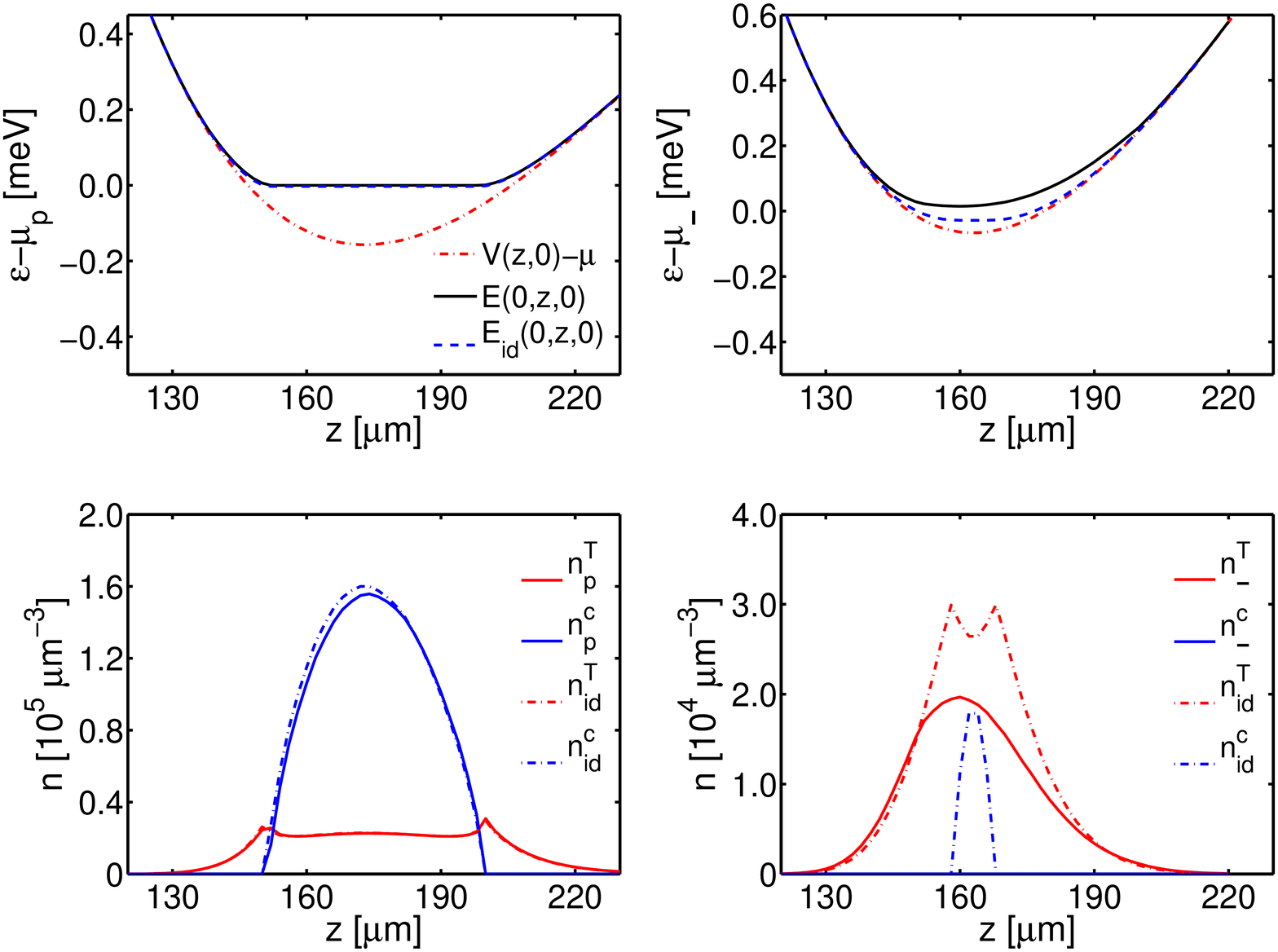}
\includegraphics[width=0.95\linewidth,bb=-190 65 739 770, clip=true]{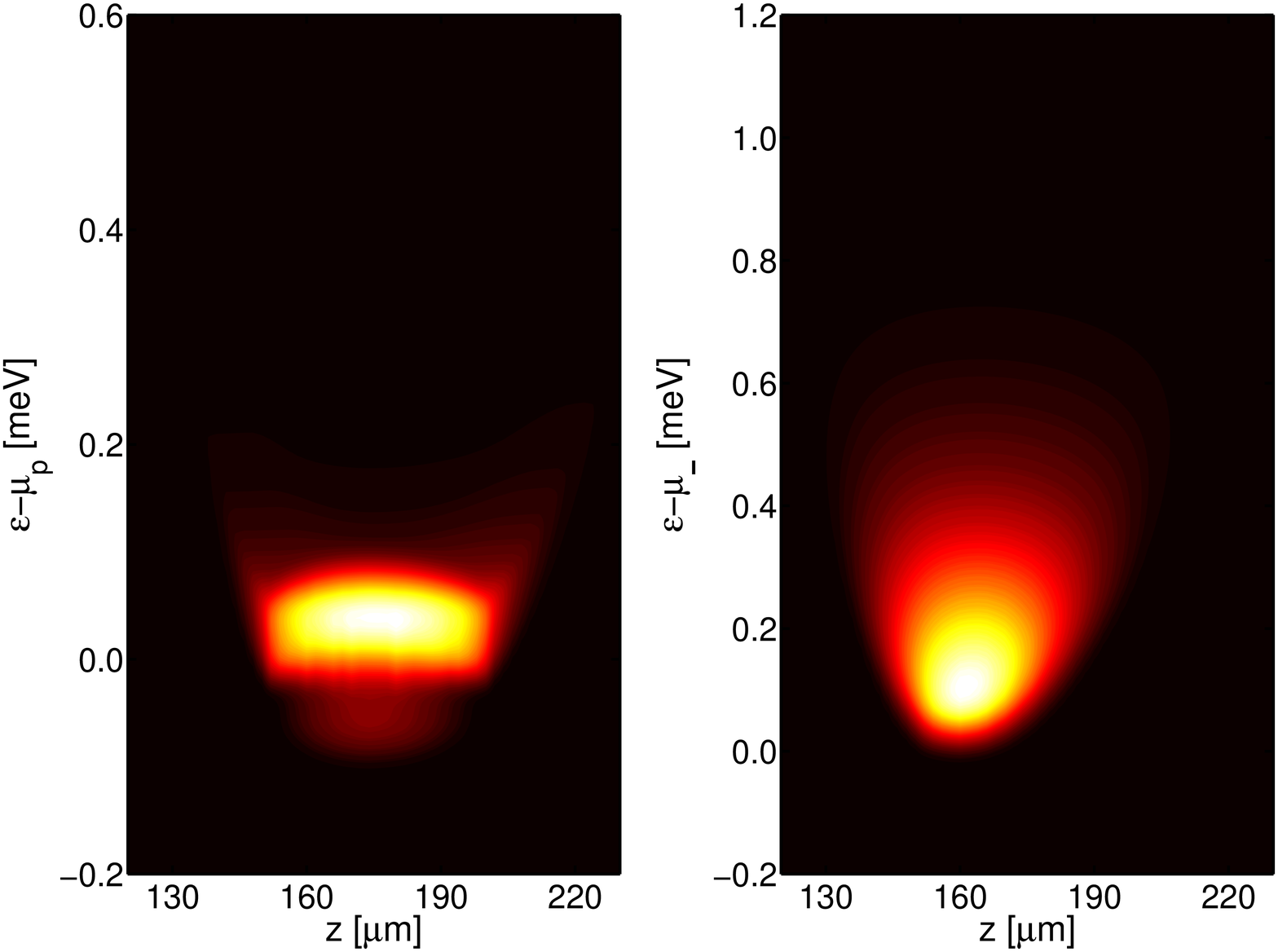}
\caption{Same presentation as in Fig.\ \protect\ref{fig1} but for a temperature of $T=\unit{1.2}{\kelvin}$ and particle numbers of $N_p=5 \times 10^9$ and $N_{-}=N_{+}=5 \times 10^{8}$ in the trap. The chemical potentials are $\mu_p=\unit{-2140}{\micro\electronvolt}$ and $\mu_{-}=\mu_{+}=\unit{-5610}{\micro\electronvolt}$.
In the middle row, additionally the densities of condensed excitons with ($n^c$) and without interspecies interaction ($n_{id}^c$) appear.}
 \label{fig3}
\end{minipage}
\end{figure}

In the following figures we show the respective trap potentials $V_i$, minimal excitation energies $E_i({\bf k}=0,z,\varrho=0)$, and density distributions of paraexcitons and orthoexcitons versus the\ $z$-coordinate. We compare the results to the case without inter-component interaction (labeled `id').
\begin{figure}[hb]
\begin{minipage}[t]{\linewidth}
\includegraphics[width=0.95\linewidth,bb=-190 65 739 770, clip=true]{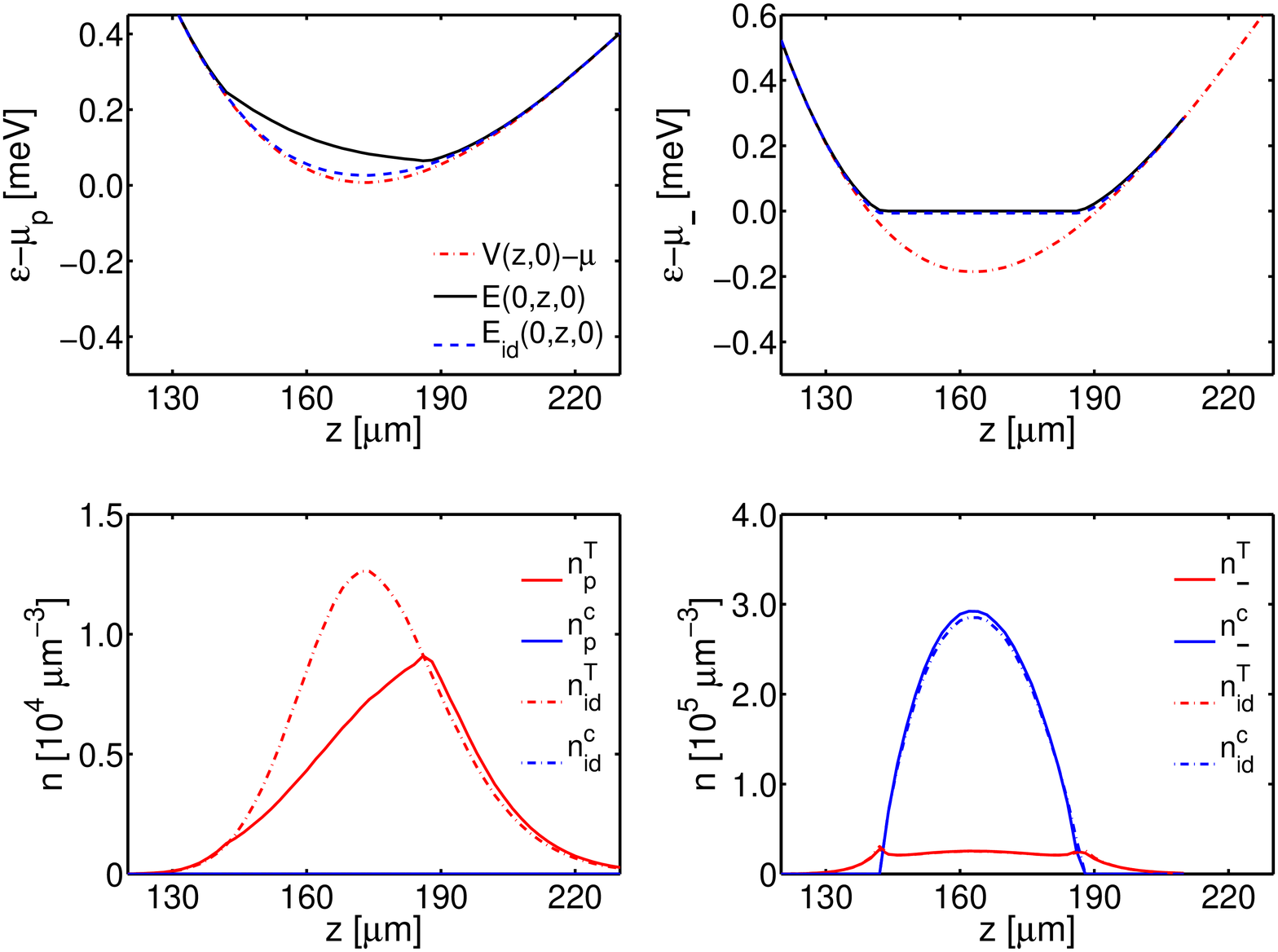}
\includegraphics[width=0.95\linewidth,bb=-190 65 739 770, clip=true]{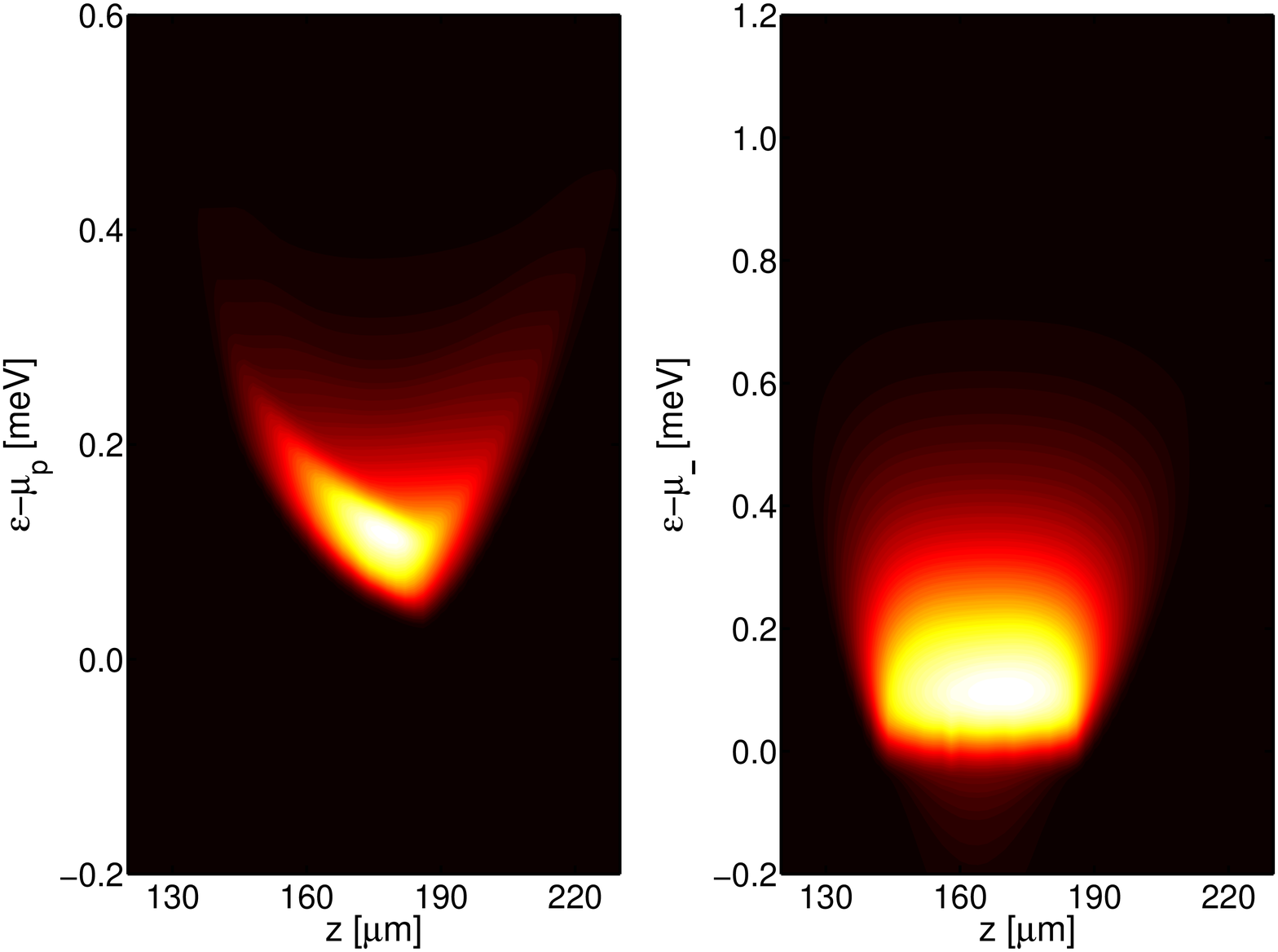}
\caption{Same presentation as in Fig.\ \protect\ref{fig1} but for a temperature of $T=\unit{1.2}{\kelvin}$ and particle numbers of $N_p=N_{+}=5 \times 10^{8}$, and $N_{-}=5 \times 10^{9}$ in the trap. The chemical potentials are $\mu_p=\unit{-2310}{\micro\electronvolt}$, $\mu_{+}=\unit{-5550}{\micro\electronvolt}$, and $\mu_{-}=\unit{-5490}{\micro\electronvolt}$.}
 \label{fig5}
\end{minipage}
\end{figure}

We first investigate case (i) by setting $N_p=5\times  10^9$, $N_\pm=5\times  10^8$ and a temperature $T=2.2$ K well above both of the estimated critical values. As Fig.\ \ref{fig1} shows, in the absence of any condensate, the lineshapes $E_i({\bf k}=0,z,\varrho=0)$ roughly follow the external potentials and the presence of multiple components causes only a weak additional renormalization. Both para- and orthoexciton densities---the latter being equal for `+' and `$-$' species---concentrate in the centers of their traps. A noticeable redistribution of the orthoexcitons with respect to the one-component case results from the large number of paraexcitons as well as from the ortho-ortho interaction. The corresponding luminescence spectra of thermal excitons are shown in Fig.\ \ref{fig1}, lower row. 
Because the modulus of the photon vector $|\mathbf{k}_0|\approx30$ \textmu m$^{-1}$ is rather small, the integrated zero-phonon spectrum of the paraexcitons almost directly resembles the minimal excitation energy $E_p({\bf k}=0,z,\varrho=0)$. In the case of orthoexcitons, every $\mathbf{k}$-vector contributes and we find a broad energy distribution above $E_\pm({\bf k}=0,z,\varrho=0)$.

Keeping the particle numbers constant, we lower the temperature to $T=1.2$ K and show case (ii) in Fig.\ \ref{fig3}.
\begin{figure}[hb]
\begin{minipage}[t]{\linewidth}
\includegraphics[width=0.95\linewidth,bb=-190 65 739 770, clip=true]{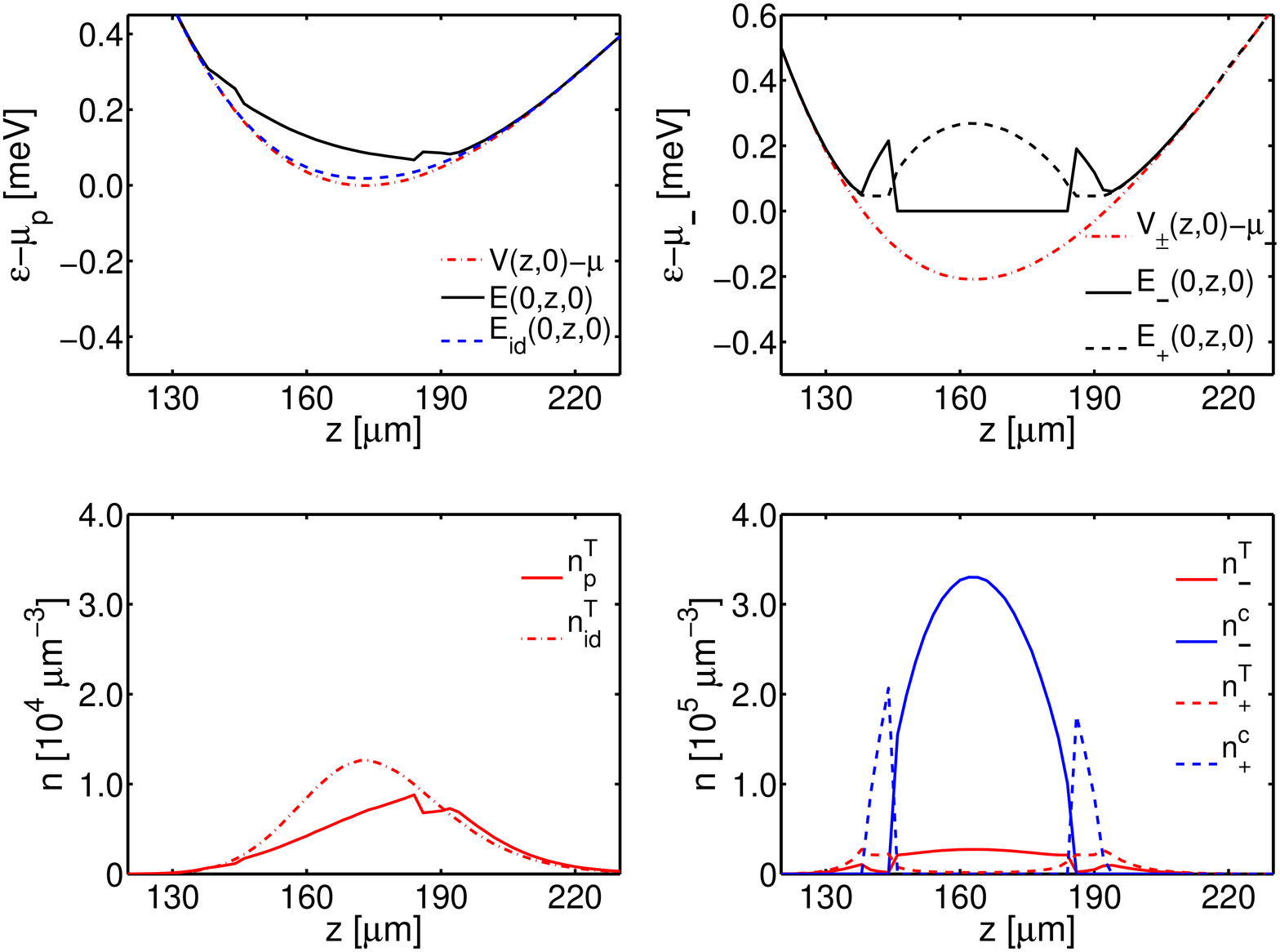}
\includegraphics[width=0.95\linewidth,bb=-190 65 739 770, clip=true]{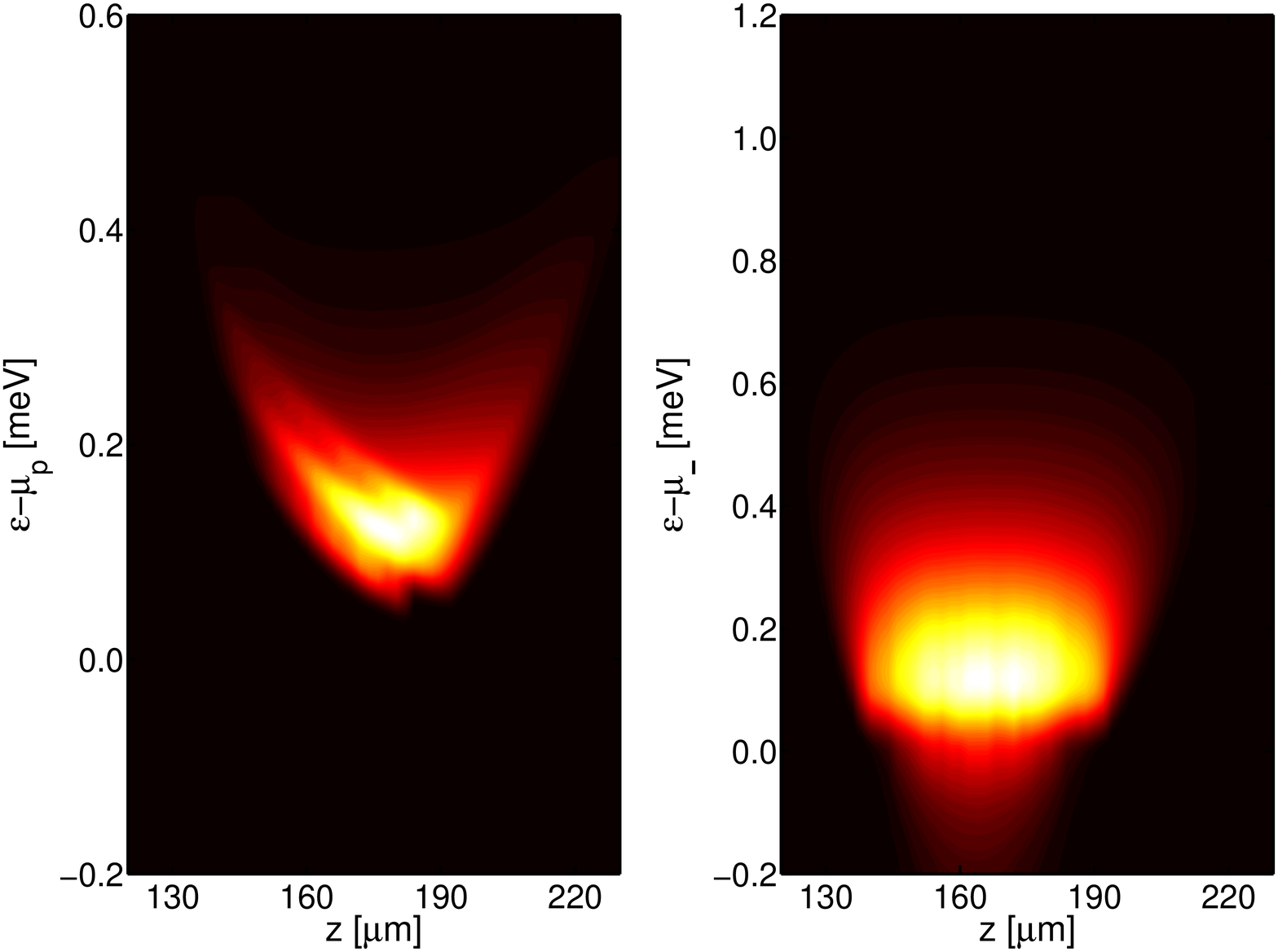}
\caption{Same presentation as in Fig.\ \protect\ref{fig1} but for a temperature of $T=\unit{1.2}{\kelvin}$ and particle numbers of $N_p=5 \times 10^8$, $N_{-}=N_{+}=5 \times 10^{9}$ in the trap. The chemical potentials are $\mu_p=\unit{-2300}{\micro\electronvolt}$, $\mu_-=\unit{-5420}{\micro\electronvolt}$, and $\mu_{+}=\unit{-5470}{\micro\electronvolt}$.}
 \label{fig7}
\end{minipage}
\end{figure}
Now the renormalized potential of the paraexcitons (Fig.\ \ref{fig3}, upper left panel) is cut at the chemical potential causing an almost flat bottom of the luminescence spectrum (Fig.\ \ref{fig3}, lower left panel). Again the densities of thermal and condensed paraexcitons show no significant deviation from the single component case (Fig.\ \ref{fig3}, middle left panel). In contrast, while isolated orthoexcitons would have been condensed, there is no ortho-condensate in the fully interacting case, which shows that the presence of multiple repulsive components lowers the critical temperature. Due to the even higher concentration of paraexcitons and the different minimum positions of the external potentials ($z_{p}=174$ \textmu m and $z_{\pm}=164$ \textmu m), the thermal orthoexcitons are slightly pushed aside (Fig.\ \ref{fig3}, middle right panel). Their spectrum (Fig.\ \ref{fig3}, lower right panel) is qualitatively nearly unchanged with respect to case (i). However, due to the lower temperature, the spectrum is less widespread. Furthermore, the chemical potential nearly touches the renormalized potential causing already a smoother curvature of the spectral shape.

If we exchange the particle numbers of paraexcitons and one species of orthoexcitons, i.e. $N_+=5 \times  10^9$, $N_p=N_-=5\times  10^8$, we realize case (iii), which is presented in Fig.\ \ref{fig5}. While in this case the density distributions of thermal and condensed ortho(+)excitons deviate only weakly from the one-component case, the displacement of the thermal paraexcitons is expressed in a heavily distorted zero-phonon spectrum.
\begin{figure}[hb]
\begin{minipage}[t]{\linewidth}
\includegraphics[width=0.95\linewidth,bb=-190 65 739 770, clip=true]{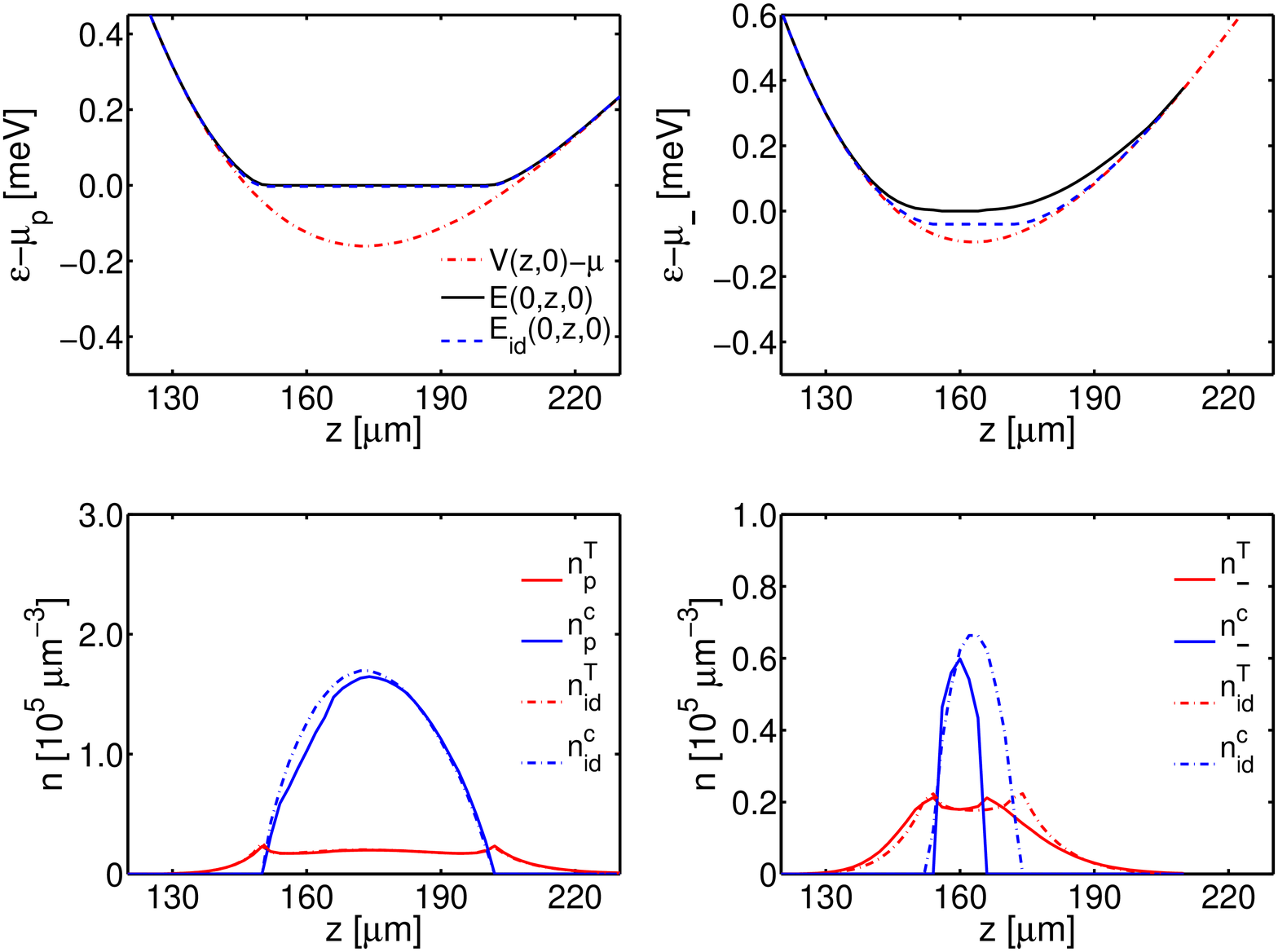}
\includegraphics[width=0.95\linewidth,bb=-190 65 739 770, clip=true]{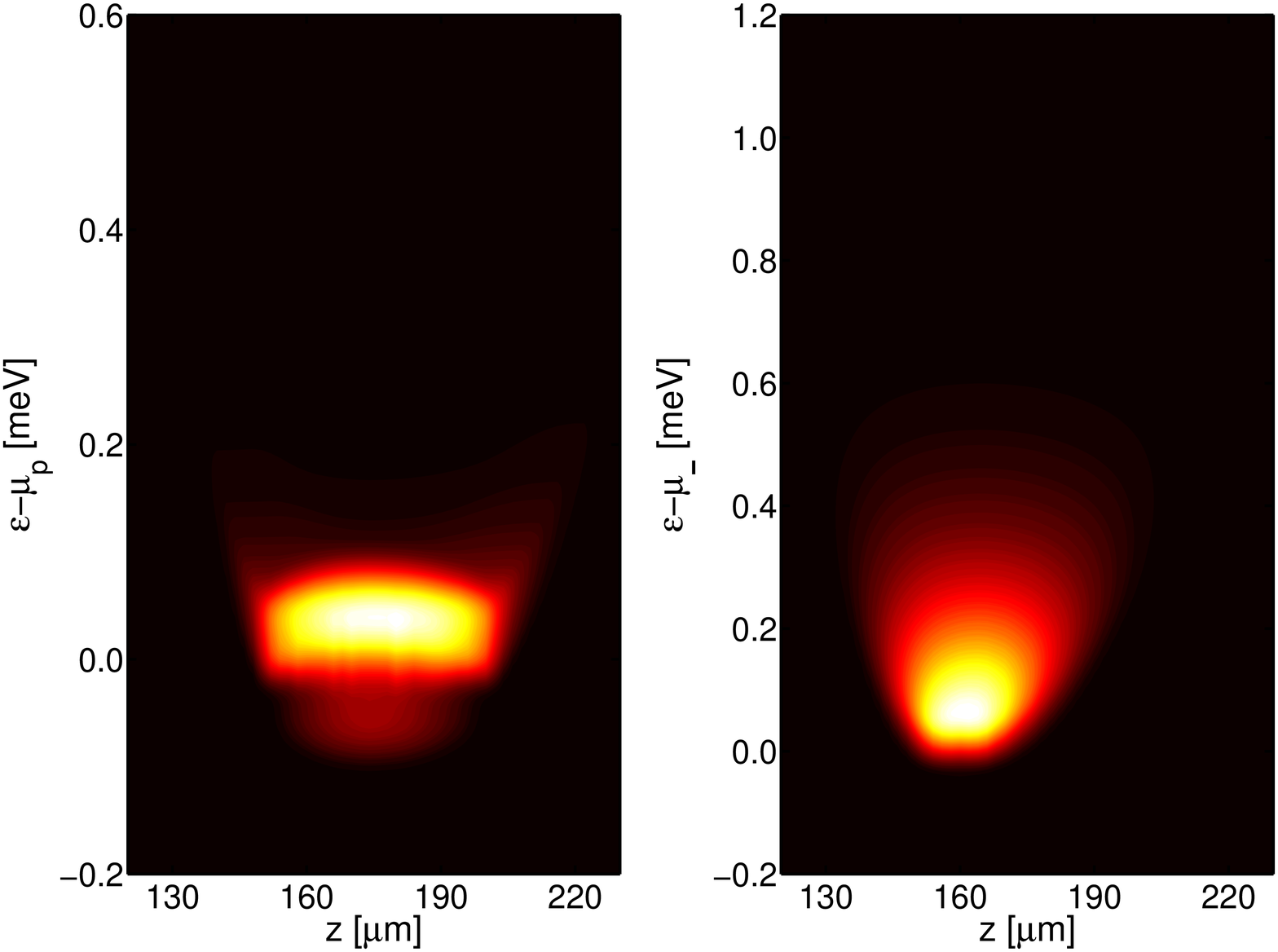}
\caption{Same presentation as in Fig.\ \protect\ref{fig1} but for a temperature of $T=\unit{1.05}{\kelvin}$ and particle numbers of $N_p=5 \times 10^9$, $N_{-}=N_{+}=5 \times 10^{8}$ in the trap. The chemical potentials are $\mu_p=\unit{-2140}{\micro\electronvolt}$, $\mu_-=\unit{-5585}{\micro\electronvolt}$, and $\mu_{+}=\unit{-5580}{\micro\electronvolt}$.}
 \label{fig9}
\end{minipage}
\end{figure}

Increasing also the particle number of the remaining ortho-species by an order of magnitude, we generate case (iv), depicted in Fig.\ \ref{fig7}.
As Shi \textit{et al.}\cite{SZC00} showed, even for finite temperature the condition for phase-separation of mutually interacting trapped condensates coincides with the $T=0$ result of Ho and Shenoy: \cite{HS96} $h_{+-}^2 > h_{++}h_{--}$. Due to their strong repulsion, the two ortho-condensates fulfill this 
condition and separate into a ball-and-shell structure with finite 
overlap, as seen in Refs.\ \onlinecite{HS96} and \onlinecite{BV97}. Yet, as found in 
Ref.\ \onlinecite{VS}, at $T>0$ no pure `+'- or `$-$'-phases exist and the 
respective thermal particles are not entirely expelled. References 
\onlinecite{BV97} and \onlinecite{OS97} pointed out that in general the component 
with the weaker self-interaction forms the outer shell. In the case of 
orthoexcitons, this criterion does not apply and the labels `+' and `$-$' 
can be interchanged in all the results presented here. In principle, 
there should exist an unstable solution with equal distributions of the 
ortho-condensates. The breaking of this symmetry is a consequence of our 
iterative numerical method. Because of the symmetry of the interaction, the paraexcitons react to the combined density of the orthoexciton species. That is why the distortion of the para-spectrum is strongest in the area of overlapping ortho-condensates.
\begin{figure}[hb]
\begin{minipage}[t]{\linewidth}
\includegraphics[width=0.95\linewidth,bb=-190 65 739 770, clip=true]{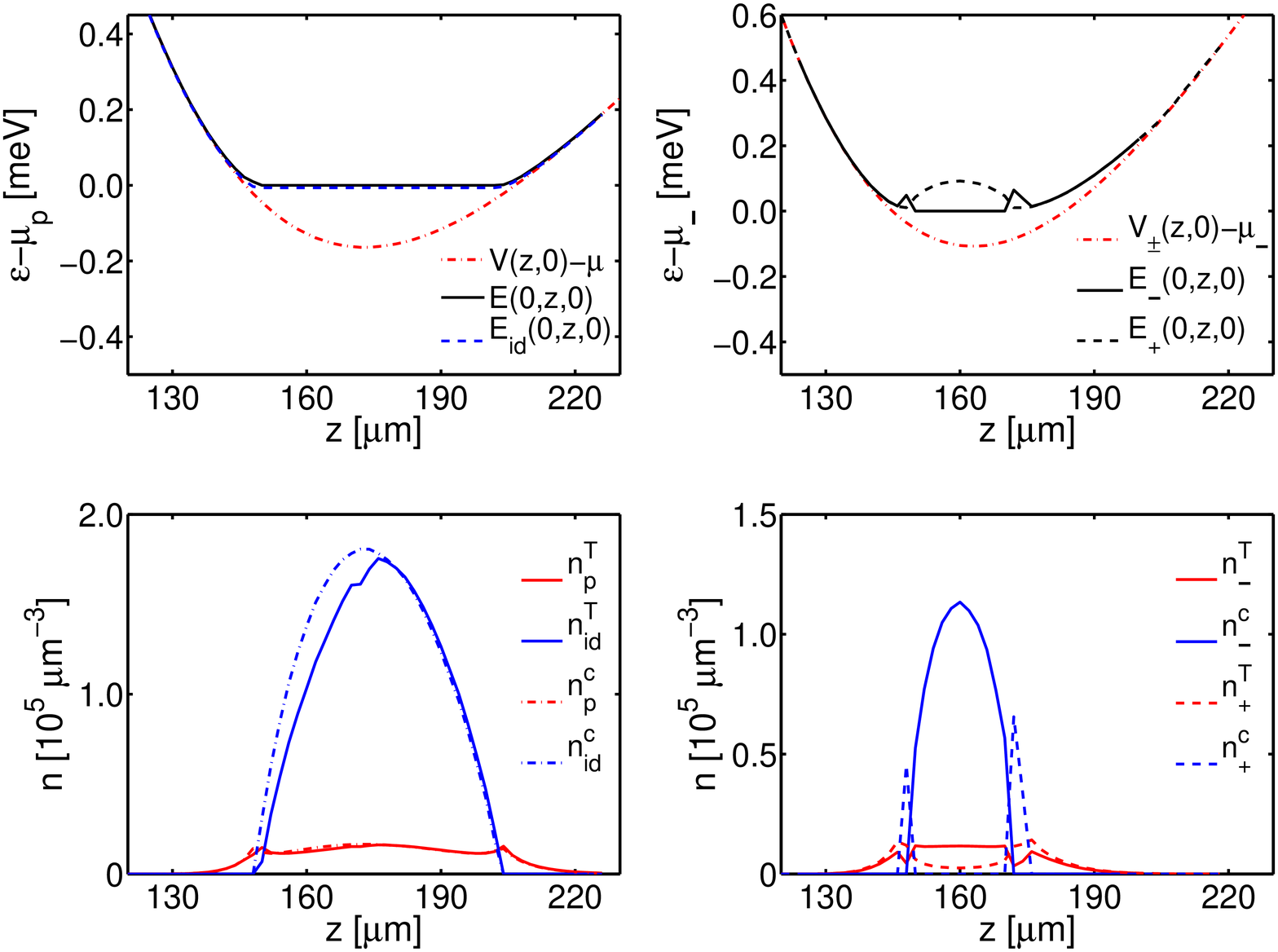}
\includegraphics[width=0.95\linewidth,bb=-190 65 739 770, clip=true]{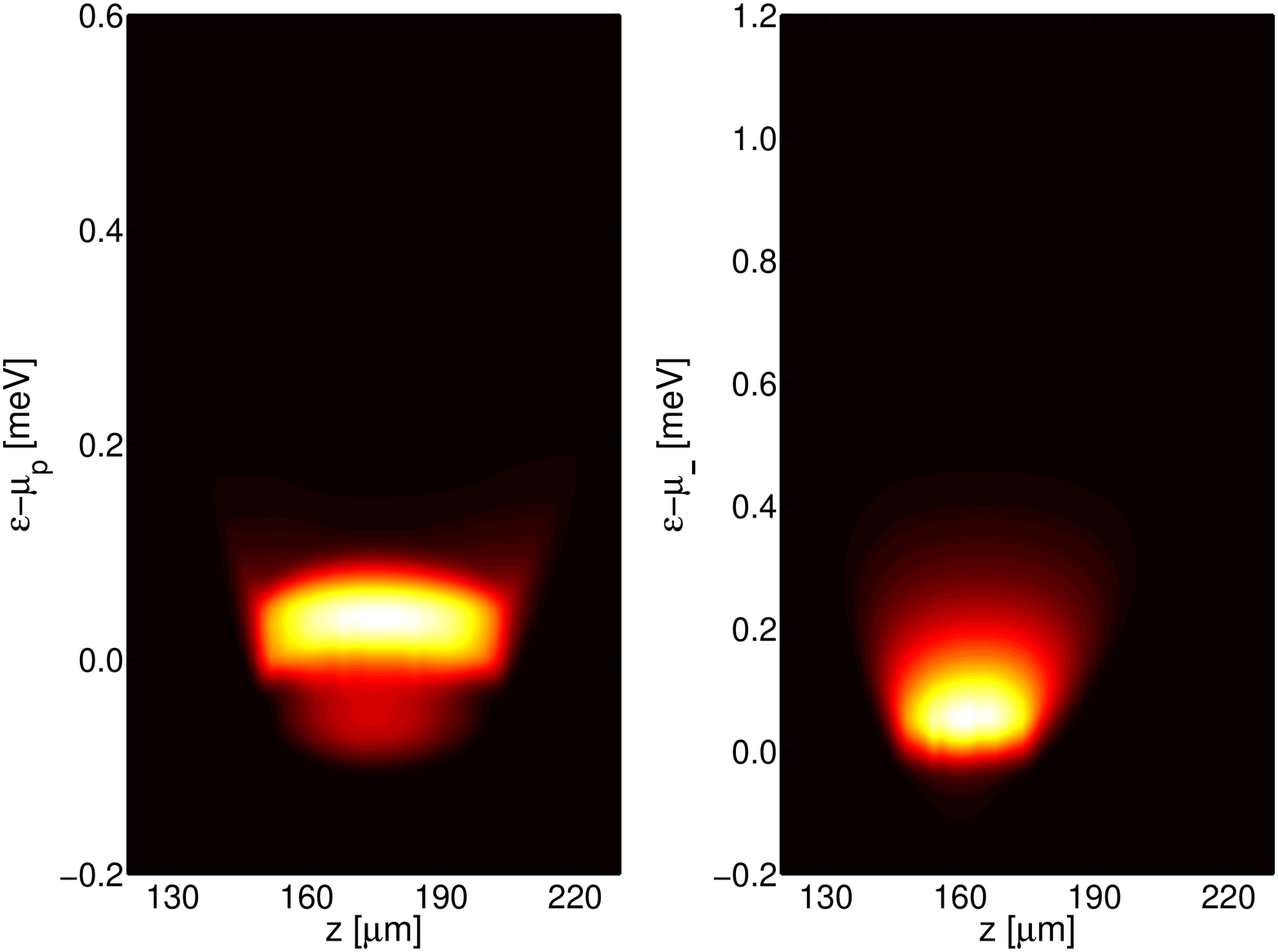}
\caption{Upper and lower row: same presentation as in Fig.\ \protect\ref{fig1}, middle row: density profiles and potentials for paraexcitons (left column) and both ortho-species (right column) for a temperature of $T=\unit{0.8}{\kelvin}$ and particle numbers of $N_p=5 \times 10^9$, $N_{-}=N_{+}=5 \times 10^{8}$ in the trap. The chemical potentials are $\mu_p=\unit{-2140}{\micro\electronvolt}$, $\mu_-=\unit{-5560}{\micro\electronvolt}$, and $\mu_{+}=\unit{-5570}{\micro\electronvolt}$.}
 \label{fig11}
\end{minipage}
\end{figure}

Let us switch back to the parameters of case (ii), i.e., $N_p=5\times 10^9$ and $N_+=N_-=5\times 10^8$, and lower the temperature to $T=1.05$ K. As Fig.\ \ref{fig9} shows, besides condensed paraexcitons we now find a small condensate of only one species of orthoexcitons (case v). While in the single-component case, for $N_+=N_-$, both ortho-species have the same critical temperature, now the mutual repulsion prevents the simultaneous condensation of the second species. We have to lower the temperature to $T=0.8$ K, to get condensates of all the components (case (vi), Fig.\ \ref{fig11}). While the ortho-species again form a ball-and-shell structure, ortho- and para-condensates do not separate because of their weak interaction ($h_{p+}^2 < h_{pp}h_{++}$).

\section{Conclusion and outlook}

We have presented a theoretical approach for the description of multi-component interacting excitonic gases in potential traps. The resulting system of equations has been subject to a number of approximations to make it numerically feasible. Finally, coupled multi-component equations for the densities of thermal excitons in Hartree--Fock--Bogoliubov--Popov approximation and the condensate densities following from the Gross--Pitaevskii equation in Thomas--Fermi approximation have been obtained and numerically solved. Compared to previous calculations, \cite{SS10} an experimentally realistic, anharmonic trap potential has been used. Six ``typical'' (but not necessarily experimentally realizable) situations leading to BEC in one or more of the species have been compared. The spatially resolved decay luminescence spectra of thermal paraexcitons and orthoexcitons exhibit clear signatures of a condensate. On the one hand, there is a flat bottom at the chemical potential known from the single-component case. \cite{SS10} On the other hand, the interspecies interaction causes additional modifications of the spectra: if there is a condensate in one of the species, the spectrum of the respective other component is clearly distorted.

In a typical experiment in the bulk or involving rather shallow potential
traps, orthoexcitons are produced by laser excitation, but are converted
fast into paraexcitons at a rate of 0.3 ns$^{-1}$. \cite{snoke,wolfe2005} Therefore,
under quasiequilibrium conditions, the particle number ratio ortho-/paraexcitons is small which corresponds to the cases (i), (ii), (v), and (vi). By increasing the stress, however, the conversion rate
decreases by more than an order of magnitude. \cite{wolfe2005} By continuous
excitation of orthoexcitons, therefore, it should be possible to obtain higher orthoexciton/paraexciton ratios like in cases (iii) and (iv).
For an exciton number of 5$\times  10^9$, the density in the center of the trap is about $10^{17}$ cm$^{-3}$ which is experimentally achievable. The same holds for the considered temperatures of $T=0.8 ... 2.2$ K. \cite{schwartz2010}

Table \ref{table1} summarizes the essential information obtained from the cases (i)--(vi) discussed above. Obviously, three conclusions can be drawn: First, if the temperature is low enough (below the respective critical temperature), every species can form a condensate. Its primary signature is a flat bottom of the respective spectrum. Second, a spatial separation occurs only between the condensates of the two ortho species, because of their strong repulsion. It shows up only in the densities, not in the combined spectrum. Therefore, it is important for future experiments to measure also the spectrally integrated density profile. Third, at occurrence of any ortho but no para condensate, the para spectrum is distorted in a characteristic way indicating a condensate in at least one of the \textit{other} species.

The presented theory is obviously only a first step towards a deeper understanding of the physics of trapped excitons. To go beyond that includes the solution of the Gross--Pitaevskii equation (without the Thomas--Fermi approximation) and the inclusion of anomalous densities already on the single-component level. Moreover, a general multi-component theory requires the consideration of mixed averages which overcomes the effective single-component picture but complicates the Bogoliubov transformation remarkably.

\begin{widetext}

\begin{table}[htb]
\centering
\begin{tabular}{*{8}{|c}|}
\hline
Case & Temperature & & & & & Orthoexciton & Paraspectrum \\
No. & (K) & $N_p/5\times 10^9$ & $N_+/5\times 10^9$ & $N_-/5\times 10^9$ & Condensate? & separation? & deformation?\\
\hline
(i) & 2.2 & 1 & 0.1 & 0.1 & - & - & -\\
(ii) & 1.2 & 1 & 0.1 & 0.1 & $p$ & - & -\\
(iii) & 1.2 & 0.1 & 0.1 & 1 & $o_-$ & - & \checkmark\\
(iv) & 1.2 & 0.1 & 1 & 1 & $o_+,o_-$ &  \checkmark &  \checkmark\\
(v) & 1.05 & 1 & 0.1 & 0.1 & $p,o_-$ & - & -\\
(vi) & 0.8 & 1 & 0.1 & 0.1 & $p,o_+,o_-$ &  \checkmark & -\\
\hline
\end{tabular}
\caption{Summary of cases (i)--(vi): occurrence of a condensate in one or more
species, of spatial separation between ortho(+) ($o_+$) and ortho($-$) ($o_-$) excitons and of a deformation of the thermal para exciton spectrum in dependence on temperature and particle numbers}\label{table1}
\end{table}

\end{widetext}

\begin{acknowledgments}
We would like to thank G.\ Manzke and F.\ Richter (Rostock), and A.\ Alvermann (Greifswald) for many fruitful discussions.
This work was supported by the Deutsche Forschungsgemeinschaft via Collaborative
Research Center SFB 652, projects B1 and B5.
\end{acknowledgments}

\end{document}